\documentstyle[aps,prl,multicol,psfig]{revtex}

\begin{document}
\draft

\title{Experimental test for subdominant superconducting phases with complex order parameters in cuprate grain boundary junctions}

\author{ W. K. Neils and D. J. Van Harlingen}
\address{Department of Physics, University of Illinois at Urbana-Champaign,
1110 W. Green St., Urbana, Illinois 61801}

\date {Received \today}
\maketitle

\begin{abstract}

We propose and implement a direct experimental test for
subdominant superconducting phases with broken time-reversal
symmetry in d-wave superconductors.  The critical current of
$45^{\circ}$-asymmetric grain boundary junctions are shown to be
extremely sensitive to the predicted onset of a complex order
parameter at (110)-surfaces and near magnetic impurities.
Measurements in YBCO and Ni-doped YBCO junctions indicate that the
symmetry at the surface is consistent with pure d-wave at all
temperatures, putting limits on the magnitude and chiral domain
structure of any subdominant symmetry component.

\end{abstract}


\begin {multicols}{2}
\narrowtext

It has recently been recognized that the order parameters of
unconventional superconductors may be unstable in the presence of
perturbations as a result of their strong magnitude and phase
anisotropy.  In the high temperature cuprate superconductors, the
$d_{x^2-y^2}$ state dominant in the bulk \cite{1,2} is readily
suppressed at surfaces, at interfaces with other materials, in
vortex cores, and in the vicinity of impurities.  This suppression
may allow the emergence of localized regions with different
symmetries, possibly including phases with complex order
parameters that break time-reversal symmetry.  These states are of
scientific interest, offering unique opportunities for studying
novel phases in superconductor systems, and may also be important
for understanding the microscopic pairing mechanism and for the
implementation of high temperature superconducting materials in
electronic devices.

The fragility of the d-wave state is a result of the phase change
of $\pi$ between orthogonal directions.  At a (110)-surface, all
specular reflection trajectories undergo a phase change of $\pi$,
forming via Andreev reflection bound surface states with zero
energy \cite{3}.  These states are observed as a zero bias
conductance peak in tunneling spectroscopy experiments
\cite{4,5,6} and as a modification of the low temperature
penetration depth \cite{7,8}.  The occupation of these states
suppresses the d-wave order parameter, and it has been suggested
that a secondary pairing interaction or subdominant component of
the primary pairing interaction could induce a complex mixture
phase at the surface \cite{9,10}.  The observation of a
zero-magnetic field splitting of the zero bias conductance peak in
a-b plane quasiparticle tunneling \cite{9,11}, may be evidence for
broken time-reversal symmetry, consistent with a complex
superconducting order parameter at the surface.  A similar effect
may occur due to scattering from magnetic impurities in the
cuprates \cite{12}, and some experiments have observed an abrupt
drop in the thermal conductivity of Ni-doped BSCCO crystals
\cite{13} that may be interpreted as the opening of an energy gap
in all directions, again consistent with the onset of a complex
order parameter.

In this Letter, we present an experiment designed to test
specifically for the formation of a complex order parameter along
the (110)-surface of YBCO films by measuring the variation of the
critical current of grain boundary junctions with temperature and
magnetic field.  We present calculations demonstrating that the
onset of a complex superconducting order parameter has a dramatic
effect on the magnitude and magnetic field diffraction patterns of
the supercurrent in this geometry.  Measurements of the critical
current in $45^{\circ}$-asymmetric grain boundary junctions of
YBCO over a wide temperature range are consistent with pure
$d_{x^2-y^2}$ symmetry, showing no evidence for complex secondary
phases.  Similar results are obtained for Ni-doped films in which
a bulk transition to a complex state has been suggested.

The most straightforward way to test for the presence of a complex
order parameter is to perform a corner SQUID or corner junction
experiment with planar faces orthogonal to the (100) and
(110)-directions \cite{1}.  In this configuration, a complex order
parameter has a phase shift between 0 and $\pi$, yielding a
characteristic modulation pattern.  However, since smooth (110)
faces are not attainable in cuprate crystals, we instead make use
of the extensively-studied bicrystal thin film grain boundary
Josephson junction.  The critical current of such junctions is a
strong function of the misorientation angle $\theta$, falling off
exponentially \cite{14} to a minimum value at $\theta=45^{\circ}$
that is roughly three orders of magnitude smaller than the thin
film critical current density.  This variation is largely due to
the due to the anisotropy of the d-wave order parameter, with
additional contributions from the structural mismatch at the
interface and band-bending effects \cite{15}.  Another important
feature of grain boundary junctions is faceting of the barrier
plane caused during film growth by the competing grain
orientations along the interface.  For most grain boundary
orientations, the faceting has little effect and the junction
critical current varies with magnetic field according to the usual
Fraunhofer diffraction pattern.  However, for
$45^{\circ}$-asymmetric junctions, in which the lobe of the bulk
d-wave order parameter in one electrode and the node in the other
are oriented normal to the interface, the critical current
modulates with applied magnetic field in a complicated manner that
reflects the detailed faceting along the grain boundary and the
symmetry of the order parameter at the junction interface
\cite{16}.

To demonstrate this, we calculate the critical current of the
grain boundary junction as a function of applied magnetic field
for different order parameter symmetries.  To model the grain
boundary junction, we assume a random distribution of facet widths
and angles.  This distribution is unique to each junction,
depending strongly on the angle and uniformity of the substrate
grain boundary interface and the growth kinetics of the deposited
cuprate film.  From AFM images, it is estimated that typical facet
widths range from $5-100nm$, with a wide range of angles.  We
assume that the local critical current density is proportional to
the product of the magnitudes of the order parameters in the two
electrodes and the relative phase across the junction, which
includes the local difference between the quantum mechanical
phases in the electrodes resulting from the unconventional
symmetry as well as that produced by the magnetic flux threading
the barrier.  Because of faceting, these quantities depend on the
local orientation of the grain boundary interface, which selects
the tunneling direction, relative to the a-b axes of the films in
the electrodes.  We consider only the effect of the applied
magnetic field, neglecting any fields produced by the tunneling
currents.  This "short junction" limit is valid since the width of
our junctions $(5-20\mu m)$ is small compared to the Josephson
penetration depth $\lambda_J = (\Phi_0/2\pi\mu_0\tau J_c)^{1/2}$,
typically $25-50\mu m$ for our devices.

Shown in Figure 1 is the meander profile for a $10\mu m$-wide
junction with 25 facets randomly-angled from $+45^{\circ}$ to
$-45^{\circ}$ from the nominal grain boundary interface.  We
calculate the critical current for this junction as a function of
the magnetic flux threading the barrier for four pairing
symmetries:  $s$, $d_{x^2-y^2}$, and the complex mixture states
$d_{x^2-y^2}+id_{xy}$ and $d_{x^2-y^2}+is$.  For s-wave, the
magnitude and phase of the order parameter is uniform, giving the
Fraunhofer diffraction modulation pattern familiar from single
slit optical interference and typically observed for Josephson
junctions in conventional BCS superconductors.  For d-wave
symmetry, the variation of facet angles causes a corresponding
variation of the magnitude and, more importantly, the sign of the
order parameter across the junction, resulting in a substantial
suppression of the critical current and a complicated modulation
with applied field that persists out to high magnetic fields.
There are several characteristic features of the field modulation
pattern for d-wave symmetry.  First, the maximum critical current
always occurs at a finite magnetic field for which the various
facets of the junction are brought into phase. Second, despite its
complex structure, the modulation pattern is always symmetric with
respect to magnetic field polarity in the short junction limit;
including self-field effects can introduce a small polarity
asymmetry due to the inhomogeneous current flow across the
junction.  Finally, the fundamental field modulation period
observed is set by the overall width and magnetic barrier of the
junction and so is the same for all random facet distributions,
whereas the critical current at low fields may exhibit either a
peak or a dip, depending on the faceting.

\begin{figure}
\centerline{\psfig{figure=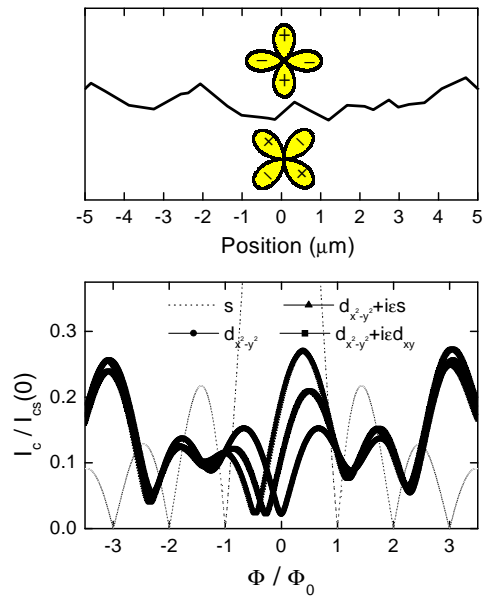,width=2.7in}} \caption{(a)
Faceting for the $45^{\circ}$-asymmetric grain boundary interface.
(b)  Calculated critical current modulation for this junction for
$s$, $d_{x^2-y^2}$, $d_{x^2-y^2}+id_{xy}$ and $d_{x^2-y^2}+is$
symmetries, assuming $\epsilon=0.20$.}
\end{figure}

For a complex order parameter, $d+id'$ or $d+is$, the modulation
pattern is significantly altered.  The zero field critical current
is dramatically enhanced, which is expected since the presence of
the secondary pairing component removes the node in the order
parameter facing the interface in one of the electrodes, allowing
a substantial tunneling contribution.  More strikingly, the
polarity symmetry is broken, resulting in an asymmetry in the
critical current modulation pattern that is is most pronounced for
small magnetic fields.  This asymmetry is a direct manifestation
of the broken time-reversal symmetry characteristic of a complex
superconducting order parameter.  For different facet
distributions, the detailed shape of the diffraction pattern
changes, but the key qualitative features remain the same.

These phenomena provide a sensitive test for the existence of a
complex order parameter in the superconducting electrodes of a
grain boundary junction.  In Figure 2(a), we demonstrate the
effect of an s-wave secondary out-of-phase order parameter
component by showing critical current diffraction patterns for
several values of $\epsilon$, the relative size of the subdominant
$s$ component in a $d_{x^2-y^2}+is$ superconductor.  To quantify
the sensitivity, Figure 2(b,c) shows the enhancement of the
zero-field critical current $I_c(0)$ and the onset of the
fractional polarity asymmetry $\alpha(H) =\{[I_c(+H)-I_c(-H)]/
[I_c(+H)+I_c(-H)]\}$, averaged over values of applied flux from
$-10\Phi_0$ to $+10\Phi_0$, as a function of $\epsilon$ for $d+is$
and $d+id'$ symmetries.  If a secondary order parameter component
onsets at a specific phase transition temperature well below
$T_c$, as has been predicted, we would expect $\epsilon$ to onset
and grow as the temperature is decreased.  Figure 2(d) shows the
critical current vs. temperature predicted for the modeled
junction, using a simple model in which $I_c(T) \sim \Delta(T)$,
which we assume to have a BCS temperature dependence.  This sharp
increase in critical current, and the simultaneous onset of
polarity asymmetry, would be a definitive signature of the
nucleation of a complex order parameter.

\begin{figure}
\centerline{\psfig{figure=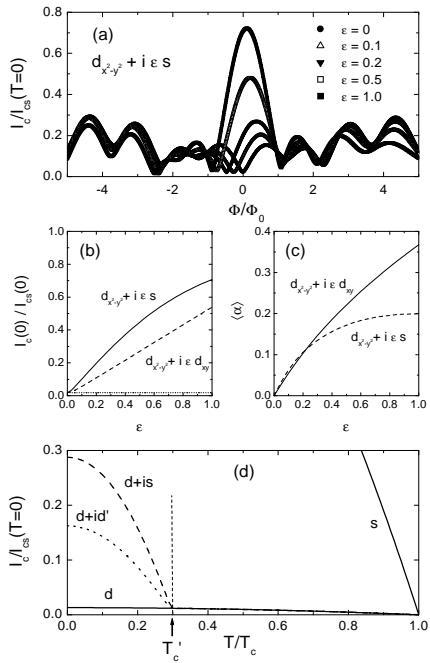,width=2.7in}} \caption{(a)
Calculated critical current modulation of the junction in Figure 1
for $d_{x^2-y^2}+is$ symmetry and a range of subdominant
components $0\leq\epsilon\leq1$.  (b)  Calculated enhancement of
$I_c(0)$ vs. $\epsilon$ for $d_{x^2-y^2}+id_{xy}$ and
$d_{x^2-y^2}+is$ symmetries.  (c)  Calculated onset of critical
current asymmetry $\alpha(H) =\{[I_c(+H)-I_c(-H)]/
[I_c(+H)+I_c(-H)]\}$ vs. $\epsilon$ for $d_{x^2-y^2}+id_{xy}$ and
$d_{x^2-y^2}+is$ symmetries.  (d)  Simulated critical current vs.
temperature for $s$, $d_{x^2-y^2}$, $d_{x^2-y^2}+id_{xy}$ and
$d_{x^2-y^2}+is$ symmetries, assuming $\epsilon=0.20$.}
\end{figure}

We have made measurements of the magnetic field variation of the
critical current of a large number of $45^{\circ}$-asymmetric
grain boundary junctions over a wide range of temperatures from
$T_c$ down to 0.3K.  YBa$_2$Cu$_3$O$_{7-x}$ thin films of
thickness 100nm were grown on $45^{\circ}$-asymmetric bicrystal
substrates by pulsed laser deposition (growth temperature 830C,
oxygen pressure 0.5Torr, power 450mJ), yielding a $T_c$ of 85-92K,
with a transition width of 1-2K, as determined from two-coil
magnetic screening measurements.  We also made films with Ni
doping ($3\%-5\%$), which had a slightly lower $T_c$ (~80K) and a
broader transition (~5K).  The films were patterned into strips of
width $5-20\mu m$ using Ar-ion milling, producing grain boundary
junctions with typical areas of $10^{-8}cm^2$.  The Ni-doped films
were measured in a dilution refrigerator down to 100mK.

To measure the critical current, we used a feedback technique in
which the bias current is automatically controlled to maintain a
small voltage level across the junction, typically $10\mu V$, as
the magnetic field or temperature is varied.  The critical
currents increase monotonically as the temperature is lowered
below $T_c$, exhibiting a nearly linear dependence over much of
the temperature range, as shown in the inset of Figure 3.  The
zero field critical current is typically $1-10\mu A$,
corresponding to an average current density of $100-1000A/cm^2$.
The modulation of the critical current at T=4K for this junction,
plotted in Figure 3, has the complex structure expected for a
junction with pure d-wave symmetry in which the order parameter
changes sign.  The critical current is maximum at about 12G,
rather than at zero magnetic field, and is nearly symmetric with
respect to polarity. The measured average asymmetry parameter
$\langle\alpha\rangle$ at the peak currents is no more than $3\%$
for this junction, typical of all junctions we have measured.

\begin{figure}
\centerline{\psfig{figure=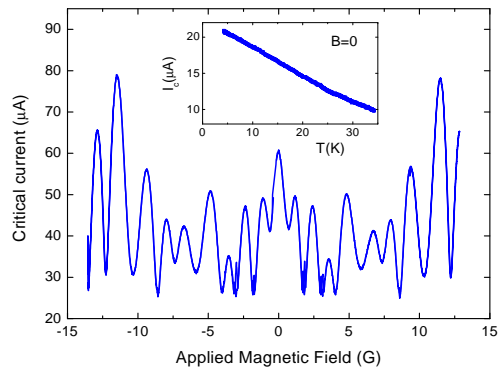,width=2.7in}} \caption{
Measured critical current modulation at $T=4K$ for a
$45^{\circ}$-asymmetric YBCO grain boundary interface. Inset shows
the measured $I_c(0)$ vs. temperature .}
\end{figure}

Figure 4 shows the critical current vs. field at a series of
different temperatures for (a) a pure YBCO, and (b) a
$3\%$-Ni-doped junction.  In both cases, the magnitude of the
critical current increases with decreasing temperature, but the
modulation pattern retains the same shape, reproducing each
detailed feature, and in particular remains symmetric with respect
to field polarity over the entire temperature range.

\begin{figure}
\centerline{\psfig{figure=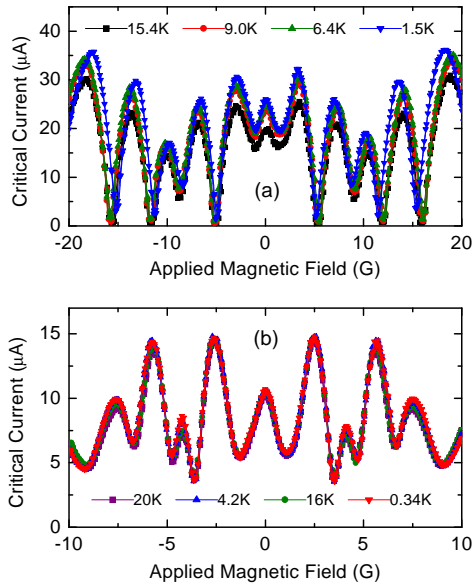,width=2.7in}}
\caption{Measured critical current modulation at a series of
temperatures for:  (a) a pure YBCO junction, and (b) a $5\%$-Ni
doped YBCO junction.}
\end{figure}

In all junctions studied to date, both pure and Ni-doped, the
polarity asymmetry of junctions cooled in zero field is typically
a few percent and is never larger than $10\%$.  This level of
asymmetry is consistent with that expected from experimental
uncertainty, self-field effects, and/or trapped magnetic flux in
the vicinity of the junction.  More significantly, we have never
observed any abrupt increases in either the zero field critical
current or the polarity asymmetry, which have been predicted to
occur at a phase transition to a state with a complex order
parameter as the junctions are cooled.  Based on a comparison of
our simulations and measurements, we would place an upper bound on
the fractional magnitude of an out-of-phase $s$ or $d_{xy}$
component added to the $d_{x^2-y^2}$ order parameter at about
$1\%$.  We note that we also do not find evidence for the onset of
a real (in-phase) subdominant order parameter component, which
would exhibit an increased zero-field critical current but no
polarity asymmetry.

It is important to consider possible reasons why we might not
observe the onset of a substantial complex order parameter, as has
been predicted by theoretical treatments.  One scenario is that
the complex order parameter region induced at the interface forms
domains of alternating chirality, e.g. $d+is$ or $d-is$.  Such
domaining is not energetically favorable due the domain wall
energy (of order the Josephson coupling energy), but could
nucleate as metastable states when a secondary phase onsets.  We
have simulated the effect of chirality domaining by putting a
random spatial distribution of domains along the interface on the
(110)-electrode of the junction.  Because of the alternating sign
of the secondary order parameter component, the net increase in
the critical current in the complex phase is substantially
reduced.  However, the domaining effectively increases the density
of facets on which the order parameter phase modulates, increasing
both the magnetic field range and the magnitude of the polarity
asymmetry.  Thus, domain formation should enhance rather than
obscure the formation of a complex order parameter at the
interface.

A second scenario is that no complex superconducting phases are
formed.  One reason could be that the relatively high transmission
coefficient of grain boundary junctions does not allow sufficient
Andreev reflection to produce the zero energy bound states that
are responsible for the suppression of the d-wave order parameter
at the surface.  However, it should be noted that the transmission
coefficient of the $45^{\circ}$-asymmetric grain boundary
junctions considered here is not particularly large, with current
densities of only $100-1000A/cm^2$.  This is supported by
measurements of the conductance vs. voltage in such junctions
which show a pronounced zero bias conductance peak \cite{17},
similar to that observed in planar tunnel junctions on thin films
and crystals with in-plane orientations.  We also note that is has
been predicted that proximity coupling between the electrodes may
induce complex phases even in high transmission junctions
\cite{18}.  A second option is that the d-wave order parameter is
suppressed, but that the conditions near the interface (density of
states, pairing interaction, ...) are not conducive to the
formation of a secondary out-of-phase superconducting component so
that the $d_{x^2-y^2}$ state remains stable at all temperatures.
In either case, the results suggest that the apparent
time-reversal symmetry breaking observed in tunneling
spectroscopy, low temperature transport, and local magnetic field
measurements may arise from some microscopic mechanism, rather
from than nucleation of a macroscopic superconducting region with
complex order parameter symmetry.  Primary candidates include
magnetic surface states, antiferromagnetic bond currents, and
barrier defects.

In conclusion, we have proposed and implemented a new experimental
test for the onset of complex order parameter symmetry at surface
and near impurities in unconventional superconductors.  The
critical current magnitude and magnetic field modulation of
$45^{\circ}$-asymmetric grain boundary junctions are shown to be
extremely sensitive to the onset of an out-of-phase secondary
order parameter, characterized by a dramatic enhancement in the
zero field current and an asymmetry with respect to magnetic field
polarity.  Measurements in YBCO and Ni-doped YBCO junctions are
consistent with pure $d_{x^2-y^2}$ symmetry, with less than $1 \%$
mixture of a subdominant superconducting phase.

We wish to thank Joe Hilliard, Chris Michael, and Tony Banks for
vital technical assistance.  This work is supported by the
National Science Foundation grant NSF-DMR99-72087 and by the
Department of Energy grant DEFDG02-96ER45439.  We acknowledge
extensive use of the Microfabrication Laboratory and the DOE
Center for Microanalysis of Materials in the Frederick Seitz
Materials Research Laboratory.

\end{multicols}
\end{document}